\documentclass[aps,prl,twocolumn,superscriptaddress,nofootinbib]{revtex4-2}

\usepackage[T1]{fontenc}
\usepackage[utf8]{inputenc}
\usepackage{amsmath,amssymb,amsfonts,amsthm}
\usepackage{bm}
\usepackage{mathtools}
\usepackage{hyperref}
\usepackage{xcolor}

\hypersetup{colorlinks=true,linkcolor=blue,citecolor=blue,urlcolor=blue}

\newcommand{\dd}{\mathrm{d}}
\newcommand{\E}{\mathbb{E}}
\newcommand{\Pp}{\mathbb{P}}

\newcommand{\one}{\mathbf{1}}

\newcommand{\C}{\mathcal{C}}

\begin{document}

\title{Critical Hawkes Processes with Random Fertilities:\\
Stationarity in Law Beyond Infinite Mean Activity}

\author{Didier Sornette}
\affiliation{Institute of Risk Analysis, Prediction and Management, Southern University of Science and Technology, Shenzhen, China}

\begin{abstract}
Genuinely critical dynamics have been proposed to organize many natural
and social systems, yet exact criticality is usually thought to preclude
stationarity because the mean activity diverges. I show that this conclusion
is not generally valid for self-exciting Hawkes point processes. At criticality,
stationarity in law is controlled not by the mean intensity, but by local
finiteness of the infinite-past Poisson-cluster construction. The relevant
object is the fixed-window hitting probability \(H_T(u)\), the probability
that a cluster born at time \(-u\) contributes at least one event to a window
of length \(T\). For memory tails \(\mathbb{P}(T>t)\sim t^{-\theta}\) and
fertility tails \(\mathbb{P}(\kappa>x)\sim x^{-\gamma}\), I prove
stationarity for \(1<\gamma<2\) and \(\theta>\gamma\) via a finite-mean-lifetime
criterion. In the finite-memory, finite-variance regime, \(H_T(u)\) is
asymptotically comparable to the cluster-survival probability, and the exact
local-finiteness condition fails. A direct asymptotic analysis of \(H_T\) gives
the sharper condition \(\theta>\gamma-1\) for stationarity to hold in the infinite-fertility-variance
regime. Thus broad fertility fluctuations can stabilize critical Hawkes
dynamics in law, producing locally finite stationary sample paths despite
infinite mean activity.
\end{abstract}

\maketitle

\paragraph{Introduction.}

Self-exciting point processes introduced by Hawkes
\cite{Hawkes1971,HawkesOakes1974} and their seismological
ETAS realizations \cite{Ogata1988,SornetteHelmstetter2002}
provide a minimal representation of cascades, aftershocks,
contagion, and social or financial excitation.  Their applications range
from collective social dynamics and online information diffusion
\cite{CraneSornette2008,Blundell2012,ZhouZhaSong2013,Zhao2015,Rizoiu2017}
to high-frequency finance and financial contagion
\cite{EmbrechtsLinigerLin2011,Bacry2013,Bacry2015,AitSahalia2015}, as well
as crime, epidemics, terrorism, and cyber risk
\cite{Mohler2011,LamprinakouGandyMcCoy2023,Jun2024,BessyRolan2021}.

In the scalar case, the central control parameter is the branching ratio
\(n\), the expected number of first-generation offspring of one event.  
The order parameter is the stationary mean intensity, which reads
\(\bar{\lambda}=\nu/(1-n)\) where \(\nu\) is the immigrant or exogenous rate.
For \(n\uparrow 1\), it diverges.  This observation
is often taken to mean that critical Hawkes processes are nonstationary or
ill-defined.  But such a conclusion uses only stationarity in expectation.
It does not decide whether the point process itself can remain locally
finite, nor whether finite-window counts or intensities can possess
time-invariant distributions with infinite first moment.  The critical
problem is therefore to determine what really fails at \(n=1\): the
existence of the process, its local finiteness, strict stationarity in law,
or only the finiteness of its mean intensity.

This is the question addressed in this paper.  Can there be a stationary
critical Hawkes process in the sense of invariant finite-window count
distributions, even though stationarity in expectation fails?  More sharply,
when the process is constructed on the whole line by starting the
immigration mechanism at time \(-\infty\), is the resulting Poisson-cluster
superposition locally finite and time-shift invariant, or does the infinite
past contribute infinitely many ancestral clusters to any bounded
observation window?

The importance of this question comes from the empirical and theoretical
status of criticality itself.  Across several domains, calibrated branching or Hawkes/ETAS-type models have
reported branching ratios close to the critical value, notably in seismicity \cite{SornetteHelmstetter2002,ZhuangWernerHarte2013},
high-frequency financial markets \cite{HardimanBercotBouchaud2013,HardimanBouchaud2014}, 
and neuronal avalanches \cite{BeggsPlenz2003,Petermann2009}.
These estimates have often been interpreted as evidence that the
underlying system is poised at, or self-organized near, a critical state.
This interpretation may of course be only asymptotic: real systems may be
slightly subcritical, while finite samples, incomplete observations,
nonstationary backgrounds, or model misspecification push estimates toward
\(n\simeq 1\).  But precisely for this reason the critical boundary cannot
be treated as a mere pathological endpoint.  It is the reference point
against which near-critical calibrations, scaling laws, and claims of
endogeneity are interpreted.

The issue is therefore to understand what the limit \(n\uparrow 1\) is a
limit toward.  If \(n=1\) necessarily destroyed the point-process
construction, then near-criticality would have to be understood only as a
subcritical approximation.  If, on the contrary, a critical Hawkes process
can remain locally finite and stationary in law while losing only
finite-mean stationarity, then empirical observations of superlinear
fluctuations, divergent sample averages, or large endogenous cascades need
not signal nonstationarity of the underlying law.  They may instead be the
observable signatures of a stationary heavy-tailed critical regime.  The
classification of critical Hawkes processes is thus essential for deciding
what is meant, mathematically and physically, when a self-exciting system is
said to operate at or near criticality.

A linear Hawkes process is exactly equivalent to an age-dependent branching,
or immigration--birth, process \cite{HawkesOakes1974}.  Immigrants arrive
according to a homogeneous Poisson process with rate $\nu$.  Each event $i$
with fertility $\kappa_i$ gives birth, independently of all other events, to
first-generation offspring according to an inhomogeneous Poisson process on
$\mathbb{R}_+$ with intensity measure $\kappa_i\phi(u)\,\dd u$, where $\phi$ is the
normalized memory kernel.  These offspring then
reproduce independently according to the same rule, with their own
fertilities.  Thus each immigrant generates a random genealogical tree, or
cluster, whose event times are measured relative to the immigrant.

The conditional intensity of the Hawkes process reads
\begin{equation}
    \lambda(t)=\nu + \sum_{t_i<t} \kappa_i\,\phi(t-t_i),
    \qquad \int_0^\infty \phi(s)\,\dd s=1,
    \label{eq:hawkes}
\end{equation}
where $\nu>0$ is the immigrant or background rate, $\phi$ is the normalized memory kernel, and $\kappa_i\geq0$ is the fertility of event $i$.  
We will consider in particular the Omori-type memory kernel
\begin{equation}
    \phi(t) \sim \frac{A}{t^{1+\theta}},\qquad t\to\infty,
    \label{eq:omori}
\end{equation}
with $\theta>0$.
 In ETAS and other heterogeneous systems, fertilities are i.i.d. random variables that are broadly distributed; we consider in particular the case
\begin{equation}
    \Pp(\kappa>x) \sim C_\kappa x^{-\gamma}, \qquad \gamma>1.
    \label{eq:fertility_tail}
\end{equation}
In the branching representation, the Hawkes branching ratio is the mean fertility,
\begin{equation}
    n = \mathbb{E}[\kappa].
\end{equation}
The standard constant-fertility Hawkes process corresponds to the degenerate case
$\kappa_i \equiv n$. The condition $\gamma>1$ guarantees the existence of the mean fertility and allows one to tune the system to criticality, $n=1$. 
By contrast, for $0<\gamma<1$, $\mathbb{E}[\kappa]$ diverges; the process then exhibits stochastic finite-time singularities \cite{SornetteHelmstetterFTS2002}.

For $n<1$, taking the statistical average of (\ref{eq:hawkes}) gives the familiar mean-rate identity 
\begin{equation}
    \bar\lambda = \nu + n\bar\lambda, ~~ \Rightarrow ~~
    \bar\lambda = \frac{\nu}{1-n}.
    \label{eq:mean_rate}
\end{equation}
This identity is the basis of the usual finite-mean stability criterion and of the conclusion that no finite stationary occurrence rate exists at $n=1$ \cite{ZhuangWernerHarte2013,BremaudMassoulie1996}.  The inference from Eq.~\eqref{eq:mean_rate} to non-existence of the process is, however, logically stronger than what the equation proves.  It proves the absence of a finite first moment at $n=1$.  It does not, by itself, rule out a stationary distribution with infinite mean.

The distinction is elementary but important.  Let $(X_j)_{j\in\mathbb{Z}}$
be independent identically distributed random variables with
$\Pp(X>x)\sim Cx^{-\mu}$ and $0<\mu<1$.  The sequence is strictly
stationary, but $\E[X_j]=\infty$.  Moreover, sums over $m$ observations
grow typically as $m^{1/\mu}$ because they are controlled by extremes,
as in the generalized central-limit/stable-law mechanism for broad
distributions \cite{BouchaudGeorges1990,Sornette2006}.  Superlinear
scaling of aggregates is therefore not a diagnostic of non-stationarity.
It can be the signature of strict stationarity with infinite first moment.

Saichev and Sornette \cite{SaichevSornette2014} made a first step in characterizing 
the statistical properties of the realizations of the Hawkes process at criticality.
For a finite but large observation window of duration $\tau$, containing on average
$\nu\tau$ immigrant sources, and assuming that no sources occurred before the
beginning of the window, Saichev and Sornette showed that the typical total
number of descendants, summed over all generations, scales at criticality $n=1$ as
\begin{equation}
    r_{\rm typ}(\nu\tau) \sim
    \begin{cases}
        (\nu\tau)^\gamma, & 1<\gamma<2,\\
        (\nu\tau)^2, & \gamma>2 \text{ or finite fertility variance},
    \end{cases}
    \label{eq:typical_scaling}
\end{equation}
while the corresponding mean is dominated by extremely rare realizations and diverges.  This proves that finite-source, finite-window critical clusters are meaningful random objects with proper distributions even when first moments fail.  
However, this result (\ref{eq:typical_scaling}) \cite{SaichevSornette2014} does not automatically ensure the existence of an infinite-past stationary Hawkes process at $n=1$, which is the focus of the present study.


\paragraph{Condition of stationarity.}

Throughout, a candidate stationary Hawkes process is understood as a
whole-line process, constructed from an immigrant process defined on the whole
time axis \(\mathbb R\), equivalently from immigration started at time
\(-\infty\).  Since each cluster evolves only as a function of its age relative
to its immigrant, the formal construction is automatically invariant under
time translations.  The only nontrivial question is whether this formal
superposition defines a genuine point process, namely whether every finite
observation window contains only finitely many events almost surely.

Let \((T_j)_{j\in\mathbb Z}\) be the immigrant times and let \(\C_j\) be
independent copies of the critical Hawkes cluster rooted at the origin.
Writing \(u\in\C_j\) for the age of an event in the \(j\)th cluster, the
absolute event time is \(T_j+u\) and the candidate random measure is
$N(\dd s)=
    \sum_{j\in\mathbb Z}\sum_{u\in\C_j}
    \delta_{T_j+u}(\dd s).
$
Hence, for any bounded observation window \(B\subset\mathbb R\),
\begin{equation}
    N(B)=
    \int_B N(\dd s)
    =
    \sum_{j\in\mathbb Z}\sum_{u\in\C_j}
    \one_{\{T_j+u\in B\}} .
    \label{eq:window_count_whole_line}
\end{equation}
The whole-line construction defines a genuine point process only if
\(N(B)<\infty\) almost surely for every bounded \(B\)
\cite{DaleyVereJones2003,Kallenberg2017,BremaudMassoulie1996}.

For \(B=[0,T]\), the relevant quantity is not merely whether an ancestral
cluster survives beyond the origin, but whether it contributes at least one
point to the fixed window \(B\).  Define the fixed-window hitting probability
\begin{equation}
    H_T(u):=\Pp\bigl(\C\cap [u,u+T]\neq\varnothing\bigr),
    \label{eq:HT_definition}
\end{equation}
where \(\C\subset\mathbb R_+\) denotes the set of event ages in a single
cluster rooted at the origin.  Immigrants born inside the window contribute a
finite Poisson number of clusters with mean \(\nu T\).  Immigrants born before
the window, at time \(-u\), contribute to \([0,T]\) precisely when their age
set intersects \([u,u+T]\).  Therefore the number of immigrant clusters that
contribute at least one point to \([0,T]\) is Poisson with mean
$\nu T+\nu\int_0^\infty H_T(u)\,\dd u$.
Consequently, provided that each single cluster is locally finite almost
surely, the exact local-finiteness condition for the whole-line construction is
\begin{equation}
    \int_0^\infty H_T(u)\,\dd u<\infty
    \qquad\text{for every }T<\infty .
    \label{eq:HT_exact_condition}
\end{equation}

Let $L:=\sup\{v:v\in\C\}$
be the lifetime of a single cluster.  Since
$
    \{\C\cap [u,u+T]\neq\varnothing\}\subseteq \{L>u\},
$
one has
\begin{equation}
    H_T(u)\leq Q(u):=\Pp(L>u).
    \label{eq:HT_le_Q}
\end{equation}
Thus the finite-mean lifetime condition
\begin{equation}
    \E[L]=\int_0^\infty Q(u)\,\dd u<\infty
    \label{eq:mean_lifetime_p_condition}
\end{equation}
is an exact sufficient condition for stationarity in law of the whole-line
construction.  It is not, however, a logically necessary condition in full
generality: the event \(L>u\) may be produced by an isolated very late
descendant, whereas \(H_T(u)\) requires a point in the fixed-width window
\([u,u+T]\).  The classification below therefore separates three levels of
control: (i) the rigorous stationary regime obtained from
\eqref{eq:mean_lifetime_p_condition}; (ii) the rigorous non-stationary regime
where \(H_T(u)\asymp Q(u)\); and (iii) the remaining long-memory regimes, where
\(H_T\) must be estimated directly.

\paragraph{Lifetime tail: a sufficient stationarity criterion.}

Let
\begin{equation}
    Q(t):=\Pp(L>t)\sim t^{-p},
    \qquad t\to\infty ,
    \label{eq:Q_lifetime_exponent}
\end{equation}
possibly up to slowly varying logarithmic factors.  If \(p>1\), then
\(\E[L]<\infty\), and \eqref{eq:HT_le_Q}--\eqref{eq:mean_lifetime_p_condition}
imply \eqref{eq:HT_exact_condition}.  This gives a rigorous sufficient
condition for stationarity in law.

The lifetime exponent follows from the standard generating-probability
formalism \cite{SaichevSornette2004}.  Let \(L_\kappa\) be the lifetime of a
cluster initiated by a source of fixed fertility \(\kappa\).  Conditional on
\(\kappa\), the number of first-generation offspring is Poisson with mean
\(\kappa\).  If \(R(t)\) denotes the probability that one such offspring branch,
including its waiting time and all its descendants, reaches beyond time \(t\),
then
\begin{equation}
    \Pp(L_\kappa\leq t)=\exp[-\kappa R(t)]~.
    \label{eq:P_kappa_SS_general}
\end{equation}
For fixed \(\kappa\), \(Q_\kappa(t):=\Pp(L_\kappa>t)  = 1- \exp[-\kappa R(t)]  \sim\kappa R(t)\) at large $t$'s, so the tail exponent of
\(Q\) is the same as that of \(R\).

Let \(G_1(z)=\E[z^r]\) be the generating function of the first-generation
offspring number, after averaging over fertilities.  In the mixed-Poisson
representation,
$
    G_1(z)=\E_\kappa\!\left[\exp[-\kappa(1-z)]\right].
$
At criticality, \(n=G_1'(1)=1\), and 
\begin{equation}
    V(z):=G_1(1-z)+z-1
    \sim
    \begin{cases}
        B_{\gamma} z^\gamma, & 1<\gamma<2,\\
        B z^2, & \gamma>2,
    \end{cases}
    \label{eq:V_small_z_corrected}
\end{equation}
with logarithmic corrections at \(\gamma=2\) \cite{SaichevSornette2004}.  Denoting \(P(t)=1-Q(t)\), 
the probability that the descendant cluster generated by a generic event has become
extinct before time $t$, the all-generation extinction equation is
\begin{equation}
    P(t)=G_1\!\left((\phi*P)(t)\right),
    \label{eq:P_self_consistent_G1}
\end{equation}
and
\begin{equation}
    R(t)=1-(\phi*P)(t).
    \label{eq:R_one_daughter}
\end{equation}
Combining these two relations gives
\begin{equation}
    R(t)-Q(t)=V(R(t)).
    \label{eq:R_Q_V_corrected}
\end{equation}

Let
$
    a(t):=1-\int_0^t\phi(s)\,\dd s
$
be the bare waiting-time tail.  For the Omori kernel 
\(\phi(t)\sim A t^{-1-\theta}\) (\ref{eq:omori}), one has \(a(t)\sim (A/\theta)t^{-\theta}\).
Using the small-\(s\) behavior
\begin{equation}
    1-\widehat\phi(s)\sim
    \begin{cases}
        A_\theta s^\theta, & 0<\theta<1,\\
        \bar\tau s, & \theta>1,
    \end{cases}
    \qquad s\to0 ,
    \label{eq:Phi_small_s_general}
\end{equation}
with logarithmic corrections at \(\theta=1\), the lifetime-tail balance is
\begin{equation}
    \mathcal D_\theta R(t)+V(R(t))\simeq a(t),
    \label{eq:lifetime_balance_general}
\end{equation}
where \(\mathcal D_\theta=D_t^\theta\) for \(0<\theta<1\), 
\(\mathcal D_\theta=\bar\tau\, d/dt\) for \(\theta>1\), up to constants,
and $D_t^\theta$ denotes the fractional derivative whose Laplace transform is $s^\theta$.

This balance (\ref{eq:lifetime_balance_general}) follows by combining the extinction equation with the temporal
embedding.  The branching equation gives \cite{SaichevSornette2004}
\begin{equation}
    Q(t)=1-G_1(1-R(t))=R(t)-V(R(t)),
    \label{thty3wg}
\end{equation}  
where $V(z)$ is defined by (\ref{eq:V_small_z_corrected}). Substituting
\(R(t)\) given by (\ref{eq:R_one_daughter}) as $R(t)=a(t)+(\phi*Q)(t)$ in (\ref{thty3wg}) yields
\[
    R(t)-(\phi*R)(t)=a(t)-\phi*V(R(t)).
\]
At large times, \(R-\phi*R\) is represented by the memory operator
\(\mathcal D_\theta R\), whose Laplace symbol is \(1-\widehat\phi(s)\), while
\(\phi*V(R)\simeq V(R)\) at leading order.   

This gives the lifetime exponent defined in (\ref{eq:Q_lifetime_exponent})
\begin{equation}
    p=
    \begin{cases}
        \min\!\left(\theta/\gamma,\;1/(\gamma-1)\right),
        & 1<\gamma<2,\\[2mm]
        \min\!\left(\theta/2,\;1\right),
        & \gamma>2 ,
    \end{cases}
    \label{eq:p_general_min_rule}
\end{equation}
again up to logarithmic corrections at the marginal values.  Therefore
\[
    p>1
    \quad\Longleftrightarrow\quad
    1<\gamma<2\quad\text{and}\quad \theta>\gamma .
\]
This proves stationarity in law in the regime
\begin{equation}
    1<\gamma<2,\qquad \theta>\gamma ,
    \label{eq:rigorous_stationary_finite_lifetime}
\end{equation}
via condition (\ref{eq:HT_exact_condition}) with inequality (\ref{eq:HT_le_Q})
because \(\E[L]<\infty\) for $p>1$.

\paragraph{Exact non-stationarity in the finite-memory, finite-variance regime ($\theta\geq 2, \gamma\geq 2$).}

The converse implication from \(\E[L]=\infty\) to non-stationarity is not
automatic.  It becomes valid when the fixed-window hitting probability $H_T(u)$  is
comparable to the lifetime survival.  In the finite-memory and finite-variance
critical regime,
$
    \theta\geq 2, \gamma\geq 2 ,
$
the lifetime tail is governed by the standard critical Bellman--Harris survival
mechanism \cite{Harris1963Branching,AthreyaNey1972}, not by an isolated extremely long waiting time.  Conditional on
survival to age \(u\), the cluster has an active genealogical population at
calendar time \(u\) with a non-vanishing probability of producing an event in
any fixed window \([u,u+T]\).  The Bellman--Harris result gives
$Q(u)=\mathbb P(L>u)\sim {c\over u}$.
In this finite-memory regime, survival up to age \(u\) is produced by an active
critical population near time \(u\), not by a single jump far beyond \(u\).
Consequently the conditional hitting probability
$
 \mathbb P\bigl(C\cap [u,u+T]\neq\varnothing\,\big|\,L>u\bigr)
$
stays bounded away from zero for every fixed \(T>0\) and therefore
for every fixed \(T>0\),
\begin{equation}
    H_T(u)\asymp Q(u)\asymp {1\over u},
    \qquad u\to\infty ,
    \label{eq:HT_Q_comparable_finite_variance}
\end{equation}
up to logarithmic corrections at \(\theta=2\) or \(\gamma=2\).  Hence
\[
    \int^\infty H_T(u)\,\dd u=\infty ,
\]
and the exact local-finiteness condition \eqref{eq:HT_exact_condition} fails.
Thus the standard whole-line critical Hawkes construction is not stationary in
law in this finite-memory, finite-variance regime.

\paragraph{Direct asymptotics of \(H_T\) in the remaining regimes.}

It remains to estimate \(H_T\) in regimes where neither
\(\E[L]<\infty\) nor the comparability \(H_T(u)\asymp Q(u)\) gives a complete
answer.  These are 
\[
    1<\gamma<2,\quad \theta \leq \gamma,
    \qquad\text{and}\qquad
    \gamma>2,\quad \theta<2 .
\]
The following derivation uses the same generating-probability formalism, but
with a fixed-window forcing.  It should be read as an asymptotic scaling
derivation: unlike the sufficient condition \(\E[L]<\infty\) and the
finite-memory/finite-variance comparability above, it is not used here as a
fully rigorous necessity proof.

Let \(R_T(u)\) be the probability that one first-generation offspring branch
hits \([u,u+T]\).  Such a hit occurs either because the offspring itself is born
in the window, or because she is born at time \(s<u\) and her descendant cluster
hits \([u-s,u-s+T]\).  Therefore
\begin{equation}
    R_T(u)
    =
    a_T(u)+(\phi*H_T)(u)~ ,
    \label{eq:RT_HT_convolution}
\end{equation}
where $ a_T(u):=\int_u^{u+T}\phi(s)\,\dd s$.
For fixed \(T\) and \(\phi(u)\sim A u^{-1-\theta}\),
$a_T(u)\sim A T u^{-1-\theta}$. Conditional on fertility \(\kappa\), no offspring branch hits the window with  
probability \(\exp[-\kappa R_T(u)]\), which generalises (\ref{eq:P_kappa_SS_general}).  Averaging over \(\kappa\) gives
$1-H_T(u)=G_1(1-R_T(u))$, or, using \(V\),
\begin{equation}
    H_T(u)=R_T(u)-V(R_T(u)).
    \label{eq:HT_RT_V}
\end{equation}
Since \(V(z)=o(z)\) as \(z\downarrow0\), \(H_T(u)\sim R_T(u)\) at the level of
leading powers.

Equations \eqref{eq:RT_HT_convolution} and \eqref{eq:HT_RT_V} yield the local
hitting balance
\begin{equation}
    \mathcal D_\theta R_T(u)+V(R_T(u))
    \simeq a_T(u)
    \sim A T u^{-1-\theta},
    \label{eq:HT_balance_general}
\end{equation}
with the same operator \(\mathcal D_\theta\) as in
\eqref{eq:lifetime_balance_general}.  This equation differs from the lifetime
balance only by replacing the tail forcing \(a(u)\sim u^{-\theta}\) by the
local-window forcing \(a_T(u)\sim u^{-1-\theta}\).

For \(1<\gamma<2\), \(V(z)\sim B_{\gamma} z^\gamma\).  The local-window balance
\(V(R_T)\sim a_T\) gives
\[
    R_T(u)\sim C_T u^{-(1+\theta)/\gamma}.
\]
For \(0<\theta<1\), this balance is self-consistent when the fractional
derivative term is smaller, which gives \(\theta>\gamma-1\).  If
\(\theta<\gamma-1\), the homogeneous fractional critical balance
\(D_t^\theta R_T+R_T^\gamma\simeq0\) gives
\(R_T(u)\sim C_T u^{-\theta/(\gamma-1)}\).  For \(\theta>1\), the finite-mean
clock gives the competing Bellman--Harris decay
\(u^{-1/(\gamma-1)}\).  Thus, ignoring marginal logarithmic corrections,
for $1<\gamma<2$, we have
\begin{equation}
    H_T(u)  \asymp
    \begin{cases}
        u^{-\theta/(\gamma-1)},
        & 0<\theta<\gamma-1,\\[1mm]
        u^{-1}\ \text{with log. corrections},
        & \theta=\gamma-1,\\[1mm]
        u^{-(1+\theta)/\gamma},
        & \gamma-1<\theta<1,\\[1mm]
        u^{-\min\{(1+\theta)/\gamma,\;1/(\gamma-1)\}}.
        & \theta>1,
    \end{cases}
    \label{eq:HT_asymptotic_infinite_variance}
\end{equation}
This scaling implies the heuristic sharp threshold
\begin{equation}
    \int_0^\infty H_T(u)\,\dd u<\infty
    \quad\Longleftrightarrow\quad
    \theta>\gamma-1,
    \qquad 1<\gamma<2,
    \label{eq:HT_heuristic_threshold_infinite_variance}
\end{equation}
with the borderline \(\theta=\gamma-1\) depending on logarithmic corrections.
In particular, the interval $1<\gamma<2,  \gamma-1<\theta\leq\gamma$
is the regime where the lifetime mean diverges, but the exact
\(H_T\)-criterion is expected to remain finite.

For \(\gamma>2\), \(V(z)\sim Bz^2\).  In the long-memory subcase
\(0<\theta<1\), the homogeneous fractional balance
\(D_t^\theta R_T+R_T^2\simeq0\) dominates the local forcing and gives
\begin{equation}
    H_T(u)\asymp u^{-\theta},
    \qquad 0<\theta<1,\quad \gamma>2 .
    \label{eq:HT_finite_variance_long_memory}
\end{equation}
For \(1<\theta<2\), the finite-mean clock gives the critical
Bellman--Harris contribution \(H_T(u)\asymp u^{-1}\), while the local forcing
would decay faster, as \(u^{-(1+\theta)/2}\).  Therefore, for $\gamma>2$, we have
\begin{equation}
    H_T(u)\asymp
    \begin{cases}
        u^{-\theta}, & 0<\theta<1,\\[1mm]
        u^{-1}\ , & 1\leq \theta<2.,
    \end{cases}
    \label{eq:HT_asymptotic_finite_variance_remaining}
\end{equation}
with possible logarithmic corrections to $ u^{-1}$ for $1\leq \theta<2$.
Both cases are non-integrable.  Hence the heuristic \(H_T\)-analysis agrees
with the exact finite-memory/finite-variance conclusion: critical Hawkes
processes with finite fertility variance are not stationary in law.

The resulting classification can be summarized as follows.  The entries marked
``proved'' follow either from the sufficient finite-mean lifetime condition or
from the finite-memory/finite-variance comparability
\(H_T(u)\asymp Q(u)\).  The entries marked ``heuristic'' follow from the direct
asymptotic balance \eqref{eq:HT_balance_general}.

\vskip 0.3cm
\noindent
{\tiny
\[
\begin{array}{c|c|c|c}
\text{Fertility regime} & \text{Memory regime}
& H_T(u)\ \text{tail} & \text{stationary in law?} \\
\hline
1<\gamma<2
& \theta>\gamma
& H_T(u)\leq Q(u),\ \E[L]<\infty
& \text{Yes, proved} \\[1ex]

1<\gamma<2
& \gamma-1<\theta\leq\gamma
& u^{-(1+\theta)/\gamma}, {1+\theta \over \gamma}>1
& \text{Yes, heuristic} \\[1ex]

1<\gamma<2
& \theta=\gamma-1
& u^{-1}\ \text{with logs}
& \text{Borderline, heuristic} \\[1ex]

1<\gamma<2
& 0<\theta<\gamma-1
& u^{-\theta/(\gamma-1)}
& \text{No, heuristic} \\[1ex]

\gamma\geq2
& \theta\geq2
& H_T(u)\asymp Q(u)\asymp u^{-1}
& \text{No, proved} \\[1ex]

\gamma>2
& 0<\theta<1
& u^{-\theta}
& \text{No, heuristic} \\[1ex]

\gamma>2
& 1\leq\theta<2
& u^{-1}\ \text{with logs}
& \text{No, heuristic} \\[1ex]

\gamma=2
& \theta<2
& \text{marginal finite-variance form}
& \text{No, heuristic}
\end{array}
\]
}

\paragraph{Physical interpretation.}
At \(n=1\), each immigrant initiates a critical cascade.  Criticality implies
an infinite expected total progeny, but not an infinite realized cascade: under
the standard branching construction, each cluster becomes extinct almost
surely.  The question for whole-line stationarity is therefore whether
immigrants from the infinite past contribute only finitely many events to every
bounded window.

The exact object controlling this question is \(H_T(u)\), not only the
lifetime tail \(Q(u)\).  The lifetime criterion \(\E[L]<\infty\) is a clean
and rigorous sufficient condition, and it proves stationarity for
\(1<\gamma<2\) and \(\theta>\gamma\).  Conversely, in the finite-memory and
finite-variance critical regime, survival to time \(u\) is produced by the
ordinary Bellman--Harris critical mechanism, so \(H_T(u)\asymp Q(u)\asymp
u^{-1}\).  The exact local-finiteness condition then fails, proving
non-stationarity.

The most delicate regime is the infinite-variance fertility case with
\(\theta<\gamma\).  There, \(Q(u)\) can be dominated by very late descendants,
while \(H_T(u)\) requires a hit in a fixed observation window.  Replacing the
tail forcing \(a(u)\sim u^{-\theta}\) by the local-window forcing
\(a_T(u)\sim u^{-1-\theta}\) provides the sharper threshold
\(\theta=\gamma-1\).  Thus, in the intermediate range
\(\gamma-1<\theta\leq\gamma\), the mean lifetime is infinite but the
whole-line process is expected to remain locally finite.  

For ETAS seismicity, the fertility exponent is
\(\gamma=\beta/\alpha\), where \(\alpha\) is the productivity exponent
with seismic moment and \(\beta\) is the moment-distribution exponent.
Typical values \(\beta\simeq2/3\) and \(\alpha\simeq1/2\) give
\(\gamma\simeq4/3\), hence the time-only stationarity threshold
$
    \theta>\gamma-1={\beta-\alpha\over\alpha}\simeq {1\over 3}.
$
Empirical Omori-memory exponents range from less than $1$ 
(which can be interpreted as the dressed Omori law $\sim 1/t^{1-\theta}$ 
\cite{SornetteSornette1999RenormalizationAftershocks,SornetteHelmstetter2002})
to as large as $1.5$. Thus, values \(\theta\simeq0.3-0.4\) \cite{UtsuOgataMatsuura1995},
place the temporal ETAS near the boundary of stationary critical dynamics in law. 

We can now reconcile the finite-window result \eqref{eq:typical_scaling} \cite{SaichevSornette2014} 
with the present classification.  The former is a genealogical statement about the total progeny generated by finitely many
immigrant families, mainly controlled by the fertility exponent \(\gamma\).
Whole-line stationarity is instead a calendar-time embedding problem: it asks
whether families born in the infinite past hit a bounded present-time window.
This depends on both the memory exponent \(\theta\) and the fertility exponent
\(\gamma\), and the correct object is the fixed-window hitting probability
\(H_T\).

\bibliographystyle{apsrev4-2}
\bibliography{reference}

\end{document}